\documentclass[12pt]{article} 

\usepackage{graphicx}
\usepackage{float}
\usepackage{cite}
\usepackage{amsfonts}
\usepackage{amssymb}
\usepackage{relsize}
\usepackage[compatibility=false]{caption}
\usepackage{subcaption}

\def\Fint{\rlap{$\Biggl\rfloor$}\Biggl\lceil}
\def\gtwid{\mathrel{\raise.3ex\hbox{$>$\kern-.75em\lower1ex\hbox{$\sim$}}}}
\def\ltwid{\mathrel{\raise.3ex\hbox{$<$\kern-.75em\lower1ex\hbox{$\sim$}}}}
\def\square{\kern1pt\vbox{\hrule height 1.2pt\hbox{\vrule width 1.2pt\hskip 3pt
   \vbox{\vskip 6pt}\hskip 3pt\vrule width 0.6pt}\hrule height 0.6pt}\kern1pt}

\begin{document}

\begin{titlepage}

\begin{flushright}
UFIFT-QG-23-04
\end{flushright}

\vskip 4cm

\begin{center}
{\bf Big Steve and the State of the Universe}
\end{center}

\vskip 1cm

\begin{center}
R. P. Woodard$^{\dagger}$
\end{center}

\begin{center}
\it{Department of Physics, University of Florida,\\
Gainesville, FL 32611, UNITED STATES}
\end{center}

\vspace{1cm}

\begin{center}
ABSTRACT
\end{center}
I share some reminiscences of the late Steven Weinberg. Then I discuss a topic 
in quantum field theory which he taught me: the role of state wave functionals in 
deriving the $i\epsilon$ term of the Feynman propagator when using the functional
formalism. This is perhaps a curiosity for in-out scattering amplitudes on flat 
space backgrounds, but it is has much greater significance for the in-in 
amplitudes of the Schwinger-Keldysh formalism in cosmology. It also touches on
the fate, about which Weinberg wondered, of the large logarithms one sometimes 
finds in quantum corrections from inflationary particle production.

\begin{flushleft}
\end{flushleft}

\vspace{2cm}

\begin{flushleft}
$^{\dagger}$ e-mail: woodard@phys.ufl.edu
\end{flushleft}

\end{titlepage}

\section{Reminiscences of Big Steve}

Physicists recall the oddest things about Steven Weinberg, often having
nothing to do with his towering scientific achievements. I first heard him
speak at a colloquium at MIT, during September of 1977. I had just met 
Nick Tsamis, a fellow freshman grad student who became Weinberg's final 
student at Harvard. Nick would also become my best friend, and a collaborator
with whom I have so far written 71 papers. Fall classes had not yet 
begun so we hiked over to nearby MIT to hear Weinberg speak on spontaneous 
symmetry breaking. My memory may have faded but I believe he was introduced 
by Viki Weisskopf, who recalled Weinberg's greatest contribution during his 
six years at MIT (1967-1973). Of course that period witnessed the paper which 
would earn the 1979 Nobel Prize, but Weisskopf passed this over to instead 
commend Weinberg's insistence that one should be able to open the windows of 
MIT's Center for Theoretical Physics! The other thing I recall from that 
colloquium is that Weinberg illustrated spontaneous symmetry breaking using 
the example of a cigar: both ends are equivalent but one end is lighted and 
the other is smoked.

Like many Harvard students, I took to calling Weinberg ``Big Steve''. He 
taught my first course on quantum field theory during the 1977-78 academic
year and it was one of the seminal experiences of my life. We students had
a saying that you must take QFT three times before really understanding it. 
In the end I took it only twice, and if I have understood the subject, it is 
due to the genius of the men who taught me: first, Steven Weinberg and then,
Sidney Coleman. Their styles were interestingly different. Whereas Sidney 
emphasized mathematical elegance, Big Steve proceeded from universal 
principles and erected a workmanlike framework of mutually reinforcing theory. 
His approach reminded me of the enormous structural strength evident in 
bridges which survive from the early industrial period of the United States.

Big Steve's QFT course was a mind-expanding experience. After most lectures
I would stumble back to my room in Perkins Hall and lay there, staring at 
the empty ceiling as my mind struggled to process the powerful concepts that 
were being impressed upon it. The course also had humorous aspects. Harvard 
policy required faculty to provide office hours, so Weinberg announced that 
his would take place right after class on Tuesdays and Thursdays. However, 
he didn't really want to sit in his office for questions which might never 
come, so each lecture ended with him apologizing that a dental appointment 
precluded holding office hours that day. That same dental appointment 
stretched throughout the entire 1977-78 academic year! The funny thing is, 
I don't recall anyone complaining: Big Steve's lectures were so brilliantly 
clear that no office hours were necessary.

History was important to Big Steve, and he made effective use of it in 
writing and in teaching. His 1972 text on Gravitation and Cosmology prefaces
its historical review of mankind's long struggle to perceive order in the
cosmos with a passage from Anna Comnena's encomium to her father, who spent
his life restoring order from the chaos which followed Manzikert. Weinberg 
devoted the first lectures of his QFT course to a historical review of 
quantum field theory, with particular attention to the many attempts at 
supplanting it. His message was that these alternatives always ended being 
recognized as either incorrect, or else consequences of quantum field theory. 
Listening to Weinberg speak made you share his conviction that QFT is the 
finest thing our species ever created. I recall an Arthur C. Clarke novel,
set more than a billion years in the future, in which a character casually
comments about needing to consult a ``field theory expert'' to understand
a particular device. You could believe in that future after hearing Weinberg
lecture.

Big Steve was full of historical anecdotes, such as Dirac's explanation 
for failing to quantize the photon field because he was afraid something 
``would go wrong''. In fact, something {\it did} go wrong when the first 
loop computations revealed ultraviolet divergences, and some physicists 
over-reacted by rejecting quantum field theory. Then came the Shelter Island 
Conference, after which the leading theorists of the day began to puzzle out 
renormalization. Weinberg recounted \cite{Shelter} how students poked fun at
the erroneous dismissal of loop corrections with the quip, ``{\it Just 
because something is infinite does not mean it is zero!}'' That story about 
divergences meant a lot to me when I later encountered skepticism about 
large logarithmic loop corrections from inflationary gravitons 
\cite{Tsamis:1996qm,Miao:2006gj,Glavan:2013jca,Wang:2014tza,Glavan:2021adm,
Tan:2021lza,Tan:2022xpn} because of the potential for gauge dependence 
\cite{Garriga:2007zk,Tsamis:2007is,Higuchi:2011vw,Miao:2011ng}. A procedure 
for removing gauge dependence is being developed \cite{Miao:2017feh,
Katuwal:2020rkv}, and I intend to channel Big Steve when announcing its success:
``{\it Just because something is gauge dependent does not mean it is zero!}'' 

Weinberg never questioned the reality of large logarithms when he discovered 
them in loop corrections to the primordial power spectrum. Quoting from the 
second of his famous papers on cosmological correlators \cite{Weinberg:2006ac}:
\begin{quotation}
In generic theories the $N$ integrals over time in $N$-th order perturbation
theory will yield correlation functions at time $t$ that grow as $\ln^N[a(t)]$. 
Such a power series in $\ln[a(t)]$ can easily add up to a time dependence that 
grows like a power of $a(t)$, or even more dramatically. As everyone knows, the
series of powers of the logarithm of energy encountered in various flat-space
theories such as quantum chromodynamics can be summed by the method of
the renormalization group. It will be interesting to see if the power series in
$\ln[a(t)]$ encountered in calculating cosmological correlation functions at 
time $t$, though arising here in a very different way, can be summed by similar
methods.
\end{quotation}
Some disagree \cite{Senatore:2009cf}, but I think Big Steve was right about the 
basic physics \cite{Kahya:2010xh}. And I hope he would appreciate the procedure 
being devised for implementing the resummation whose potential he foresaw 
\cite{Miao:2021gic,Woodard:2023rqo}.

Everyone who had the privilege of working with Weinberg knows the enthusiasm 
he brought to research, and his determination to overcome all obstacles. I 
well recall the time he asked me to visit the UT in 2005, in order to consult 
about his work on loop corrections to the primordial power spectrum 
\cite{Weinberg:2005vy}. He had just realized that the in-out amplitudes of 
conventional quantum field theory are not the appropriate objects of study 
in cosmology because they are dominated by assumptions about the yet-unknown 
future, and because even the matrix elements of Hermitian operators are not
generally real. I cannot count the number of times I have grown irritated
listening to intellectually dishonest colleagues attempt to avoid these 
problems by devising tricks to make the in-out formalism accomplish something 
for which it was never intended, and for which it is not well suited. Not Big 
Steve. He understood that true expectation values are the better object of 
study in cosmology and, not knowing of the Schwinger-Keldysh formalism 
\cite{Schwinger:1960qe,Mahanthappa:1962ex,Bakshi:1962dv,Bakshi:1963bn,
Keldysh:1964ud,Chou:1984es,Jordan:1986ug,Calzetta:1986ey}, he devised a 
Hamiltonian technique for computing them \cite{Weinberg:2005vy}, which is now 
more commonly used than Schwinger's method. Weinberg only learned of Schwinger's 
work from his then-student, Bua Chaicherdsakul, and he of course gave full 
credit to Schwinger thereafter. 

Aside from being blown away by Big Steve having invented his own version of 
the Schwinger-Keldysh formalism, the incident I most recall from that visit 
came during the morning of my final day. I had gotten up early and was waiting 
in the hotel before departing for the airport. The phone rings and who should 
it be but Steven Weinberg, apologizing for the early call and asking to discuss 
a physics issue which had been troubling him. Another passage from that Clarke 
novel comes to mind, about the men of the Dawn Ages never permitting problems 
to hold them up for very long.

I'll close by recalling part of the final e-mail notes we exchanged. I had 
written to congratulate him on winning the Breakthrough Prize, and to share 
some sad news about a mutual friend. That was September of 2020, when the
pandemic still dominated our lives, and I mentioned the travel bans which had
kept my wife and me apart for six months. Weinberg replied:
\begin{quotation}
I am glad that you and your wife are together again. My wife and I are together, 
fanatically isolated at home, but both of us getting a lot of work done and 
staying safe.
\end{quotation}
He was 87 at the time! I recall thinking how fine it would be, should I chance 
to reach that age, to remain so active and so passionate about 
physics.\footnote{I count {\bf seventeen} books and papers written after he 
had turned 80 \cite{Weinberg:2013vgl,Weinberg:2013cfa,Weinberg:2013kea,
Weinberg:2014qru,Weinberg:2014ewa,Weinberg:2016kyd,Weinberg:2016axv,
Weinberg:2016uml,Flauger:2017ged,Weinberg:2018apv,Weinberg:2019mai,
Flauger:2019cam,Weinberg:2020zba,Bousso:2020ukx,Weinberg:2020nsn,
Weinberg:2021exr,Weinberg:2021kzu}!} Big Steve was a force of nature; it was 
a privilege to have known him.

\section{The True Origin of the $i\epsilon$}

Weinberg believed that everything about quantum field theory should be derived.
He used the term ``folk theorems'' to describe commonly accepted beliefs for
which no derivation was currently available. I have chosen the topic for this
article to be a minor but irritating lacuna in the derivation of propagators
from the functional formalism. It was a point Big Steve derived for us back in 
1977-78, which is usually resolved by hand-waving.\footnote{A Russian colleague 
pointed out a similar derivation by Slavnov and Faddeev \cite{Faddeev:1980be}.} 
I refer to the famous ``$i \epsilon$'' part of the Feynman propagator. To make 
the discussion transparent I will work in the context of a Simple Harmonic 
Oscillator whose position is $q(t)$, and whose dynamics are controlled by the 
Lagrangian,
\begin{equation}
L = \frac12 m \dot{q}^2 - \frac12 m \omega^2 q^2 \; . \label{SHOLag}
\end{equation}

The propagator we will use the functional integral formalism to derive is,
\begin{eqnarray}
\lefteqn{i\Delta(t;t') \equiv \Bigl\langle \Omega \Bigl\vert T\Bigl[q(t) 
q(t')\Bigr]\Bigr\vert \Omega \Bigr\rangle = \frac{\hbar \, e^{-i \omega \vert 
t - t' \vert}}{2 m \omega} \; , } \label{prop1} \\
& & \hspace{-0.5cm} = \!\!\lim_{\epsilon \rightarrow 0^+} \!\frac{\hbar}{m} \!\!
\int_{-\infty}^{\infty} \!\! \frac{d k^0}{2\pi} \frac{i e^{-i k^0 (t - t')}}{
(k^0 \!-\! \omega \!+\! i \epsilon) (k^0 \!+\! \omega \!-\! i \epsilon)} 
= \!\!\lim_{\epsilon \rightarrow 0^+} \!\frac{\hbar}{m} \!\!\int_{-\infty}^{\infty} 
\!\! \frac{d k^0}{2\pi} \frac{i e^{-i k^0 (t - t')}}{(k^0)^2 \!-\! \omega^2 \!+\! 
i \epsilon} . \label{prop2} \qquad
\end{eqnarray}
Of course there is absolutely no ambiguity about the canonical derivation, 
which proceeds from (\ref{prop1}) to (\ref{prop2}). The issue is how we would 
get (\ref{prop2}) directly from the functional integral formalism,
\begin{equation}
\Fint [dq] \, q(t) q(t') \exp\Biggl[\frac{i}{\hbar} \int_{-\infty}^{\infty} 
\!\!\!\!\!\! ds \Bigl\{\frac12 m \dot{q}^2(s) - \frac12 m \omega^2 q^2(s)\Bigr\} 
\Biggr] \; . \label{Fint}
\end{equation}
The short answer is that it just isn't possible. Although expression (\ref{Fint})
should result in {\it some} function which obeys the propagator equation,
\begin{equation}
\frac{i m}{\hbar} \Bigl( \frac{d^2}{dt^2} + \omega^2\Bigr) i\Delta(t;t') = 
\delta(t \!-\! t') \; , \label{propeqn}
\end{equation}
it is not at all clear how $i\Delta(t;t')$ acquires the correct real part from 
an exponent which is entirely imaginary. Some QFT texts argue that the $i 
\epsilon$ follows from complex analysis, by the need to deform the functional 
integration over $q(s)$ into the complex plane. However, this is problematic 
when one considers the four different propagators of the Schwinger-Keldysh 
formalism which each follow from a functional integral. Other texts argue 
that the $i \epsilon$ has something to do with the temporal integration 
running from $-\infty$ to $+\infty$, but this is also problematic when one 
considers that the free Lagrangian (\ref{SHOLag}) must produce the very same 
propagator (\ref{prop2}), no matter when the initial and final states are 
specified. Big Steve had no patience with this sort of hand-waiving, and he 
discovered a better explanation.

The key is incorporating the ground state wave function in the functional 
integral expression (\ref{Fint}). For the simple harmonic oscillator (\ref{SHOLag}) 
this is,
\begin{equation}
\Omega(q) = \Bigl[\frac{\pi m \omega}{\hbar}\Bigr]^{\frac14} \exp\Bigl[-
\frac{m \omega}{2 \hbar} q^2\Bigr] \; . \label{ground}
\end{equation}
Let us also make the range of temporal integration finite, and employ a 
classical source $J(s)$,
\begin{eqnarray}
\lefteqn{Z[J] \equiv \Fint [dq] \, 
\Omega^*\Bigl(q(t_f)\Bigr) } \nonumber \\
& & \hspace{1cm} \times \exp\Biggl[\frac{i}{\hbar} \int_{t_i}^{t_f} \!\!\!\!\!\! 
ds \Bigl\{\frac12 m \dot{q}^2(s) - \frac12 m \omega^2 q^2(s) + J(s) q(s)\Bigr\} \Biggr] 
\Omega\Bigl(q(t_i)\Bigr) \; . \qquad \label{Zdef}
\end{eqnarray}
Note that the functional integration in expression (\ref{Zdef}) only extends
over functions $q(s)$ for $t_i \leq s \leq t_f$. We should be able to recover the
propagator (\ref{prop2}) by functionally differentiating $Z[J]$,
\begin{equation}
i\Delta(t;t') = \frac{(-i\hbar)^2 \delta^2 Z[J]}{\delta J(t) \delta J(t')} 
\Biggl\vert_{J = 0} \; . \label{putative}
\end{equation}

Expression (\ref{Zdef}) is the functional integral of $\exp[\frac{i}{\hbar} E[q,J]]$,
with exponent,
\begin{equation}
E[q,J] = \frac{i m \omega}{2} q^2(t_f) + \int_{t_i}^{t_f} \!\!\!\! ds \Bigl[\frac12
m \dot{q}^2(s) - \frac12 m \omega^2 q^2(s) + J(s) q(s)\Bigr] + \frac{i m \omega}{2}
q^2(t_i) \; . \label{exponent}
\end{equation}
The result of any Gaussian integral is the exponential evaluated at its stationary 
point. To find this stationary point we vary (\ref{exponent}), taking care to 
include the surface variations,
\begin{eqnarray}
\lefteqn{ \frac{\delta E[q,J]}{\delta q(t)} = m \delta(t \!-\! t_f) 
\Bigl[\dot{q}(t_f) \!+\! i \omega q(t_f)\Bigr] } \nonumber \\
& & \hspace{2.5cm} - m \Bigl[ \ddot{q}(t) \!+\! \omega^2 q(t) \!-\! \frac{J(t)}{m}
\Bigr] - m \delta(t \!-\! t_i) \Bigl[\dot{q}(t_i) \!-\! i \omega q(t_i)\Bigr] \; .
\qquad \label{variation}
\end{eqnarray}
Note that the surface variations enforce Feynman boundary conditions, which is the
part one couldn't get without including the initial and final states. Setting
(\ref{variation}) to zero has the unique solution,
\begin{equation}
q[J](t) = \frac{i}{\hbar} \int_{t_i}^{t_f} \!\!\!\! dt' \, i\Delta(t;t') J(t') 
\; . \label{qJ}
\end{equation}
Partial integration makes it simple to evaluate the exponent (\ref{exponent}) at 
its stationary point,
\begin{eqnarray}
\lefteqn{ E\Bigl[ q[J],J\Bigr] = \frac{i m}{2} q(t_f) \Bigl[\dot{q}(t_f) \!+\!
i \omega q(t_f)\Bigr] + \int_{t_i}^{t_f} \!\!\!\! dt \Biggl\{ -\frac12 m q(t) 
\Bigl[\ddot{q} \!+\! \omega^2 q(t) \!-\! \frac{J(t)}{m}\Bigr] } \nonumber \\
& & \hspace{4.5cm} + \frac12 q(t) J(t) \Biggr\} + \frac{i m}{2} q(t_i) \Bigl[
\dot{q}(t_i) \!-\! i \omega q(t_i)\Bigr] \; , \qquad \\
& = & \frac{i}{2 \hbar} \int_{t_i}^{t_f} \!\!\!\! dt J(t) \!\! \int_{t_i}^{t_f}
\!\!\!\! dt' J(t') \, i\Delta(t;t') \; . \label{EofJ}
\end{eqnarray}
Hence we conclude,
\begin{equation}
\ln\Bigl[Z[J]\Bigr] = -\frac{1}{2 \hbar^2} \int_{t_i}^{t_f} \!\!\!\! dt J(t) 
\!\! \int_{t_i}^{t_f} \!\!\!\! dt' J(t') \, i\Delta(t;t') \; ,
\end{equation}
which obviously gives the Feynman propagator in expression (\ref{putative}).

The generalization to quantum field theory is straightforward; in making it
I shall adopt the usual convention of setting $\hbar = c = 1$. The Lagrangian
density of a free scalar field on flat space with spacelike signature is,
\begin{equation}
\mathcal{L} = -\frac12 \partial_{\mu} \varphi \partial^{\mu} \varphi - \frac12
m^2 \varphi^2 \; . \label{scalarL}
\end{equation}
It seems that application of quantum field theory brings me back to something
else Big Steve taught us: that all free theories become simple harmonic 
oscillators in spatial Fourier coordinates,
\begin{equation}
\widetilde{\varphi}(t,\vec{k}) \equiv \int \!\! d^3x \, e^{i \vec{k} \cdot \vec{x}}
\varphi(t,\vec{x}) \; .
\end{equation}
We can identify each modes' mass and frequency by using Parseval's theorem
on the Lagrangian,
\begin{equation}
L \equiv \int \!\! d^3x \mathcal{L} = \int \!\! \frac{d^3k}{(2\pi)^3} \Bigl\{
\frac12 \dot{\widetilde{\varphi}}(t,\vec{k}) \dot{\widetilde{\varphi}}^*(t,\vec{k})
+ \frac12 (k^2 \!+\! m^2) \widetilde{\varphi}(t,\vec{k}) 
\widetilde{\varphi}^*(t,\vec{k}) \Bigr\} \; .
\end{equation}
Comparison with (\ref{SHOLag}) implies the identifications,
\begin{equation}
\omega \rightarrow \sqrt{k^2 + m^2} \qquad , \qquad m \longrightarrow 
\frac{d^3k}{(2\pi)^3} \; . \label{indent}
\end{equation}
Hence the ground state wave functional is,
\begin{eqnarray}
\Omega\Bigl[\varphi(t)\Bigr] &\!\!\! = \!\!\!& N \exp\Biggl[ \frac12 \sum_{\vec{k}}
m \omega \, \widetilde{\varphi}(t,\vec{k}) \widetilde{\varphi}^*(t,\vec{k}) \Biggr]
\; , \\
&\!\!\! = \!\!\!& N \exp\Biggl[ \frac12 \int \!\! \frac{d^3k}{(2\pi)^3} \, 
\widetilde{\varphi}(t,\vec{k}) \sqrt{k^2 + m^2} \, \widetilde{\varphi}^*(t,\vec{k})
\Biggr] \; , \\
&\!\!\! = \!\!\!& N \exp\Biggl[ \frac12 \int \!\! d^3x \, \varphi(t,\vec{x}) 
\sqrt{-\nabla^2 + m^2} \, \varphi(t,\vec{x}) \Biggr] \; . \label{groundphi}
\end{eqnarray}  

\section{State Wave Functionals in Cosmology}

I hope people will agree that Big Steve's functional derivation of the $i 
\epsilon$ in the propagator is far superior to the usual hand-waiving. 
However, we already knew the answer from the canonical formalism. The real
efficacy of state wave functionals becomes evident when applying the 
Schwinger-Keldysh formalism to study the time evolution of expectation 
values in cosmology.

\subsection{Cosmological Particle Production}

Quantum field theory in cosmology is much more interesting than its flat space
cousin owing to what Schr\"odinger termed ``the alarming phenomena'' of particle
production ``caused by accelerated expansion.'' \cite{Schrodinger:1939,Parker:1968mv,
Parker:1969au,Parker:1971pt}. To understand this, consider a homogeneous, isotropic 
and spatially flat geometry with scale factor $a(t)$, Hubble parameter $H(t)$ and 
first slow roll parameter $\epsilon(t)$,
\begin{equation}
ds^2 = -dt^2 + a^2(t) d\vec{x} \!\cdot\! d\vec{x} \qquad \Longrightarrow \qquad
H(t) \equiv \frac{\dot{a}}{a} \quad , \quad \epsilon(t) \equiv -\frac{H}{H^2} \; .
\label{geometry}
\end{equation}
Students of QFT are familiar with how the Energy-Time Uncertainty Principle 
permits a virtual particle of energy $E = \sqrt{k^2 + m^2}$ to exist for a time 
$\Delta t \ltwid 1/E$. The simplest way of understanding many QFT effects is by
positing the existence of these virtual particles and then using classical field
theory, from which it follows that the strongest effects come from virtual
particles with the longest persistence times $\Delta t$. An expanding universe 
($H > 0$) strengthens QFT effects because the momentum $k$ redshifts to increase
the persistence time, which is governed by the integral,
\begin{equation}
\int_{t}^{t + \Delta t} \!\!\!\!\!\!\!\!\! dt' \sqrt{ \frac{k^2}{a^2(t')} + m^2} 
\qquad \ltwid \qquad 1 \; . \label{ETUR}
\end{equation}
Just as in flat space, the longest-lived virtual particles are the lightest.
Taking the massless limit gives,
\begin{equation}
\int_{t}^{t + \Delta t} \!\!\!\!\!\!\!\!\! dt' \frac{k}{a(t')} \qquad \ltwid \qquad
1 \; . \label{massless}
\end{equation}
For accelerated expansion ($\epsilon < 1$) the integral converges even as 
$\Delta t \rightarrow \infty$. Hence a sufficiently long wave length virtual 
particle can live forever.

Although correct, the preceding discussion fails to account for the rate $dN/dt$
at which virtual particles emerge from the vacuum. This is very important because
almost all massless particles possess a killer symmetry known as conformal 
invariance which means they cannot tell the difference between a metric 
$g_{\mu\nu}(x)$ and another $\Omega^2(x) \times g_{\mu\nu}(x)$. Changing the time 
coordinate $t$ of the cosmological geometry (\ref{geometry}) to ``conformal time''
$\eta$ with $d\eta \equiv dt/a(t)$ makes the geometry conformal to flat space,
\begin{equation}
ds^2 = -dt^2 + a^2 d\vec{x} \!\cdot\! d\vec{x} = a^2 \Bigl[-d\eta^2 + d\vec{x}
\!\times\! d\vec{x}\Bigr] \; . \label{conformal}
\end{equation}
This means that the emergence rate in conformal coordinates is the same as in
flat space,
\begin{equation}
\frac{dN}{d\eta} = \Gamma_{\rm flat} \qquad \Longrightarrow \qquad \frac{dN}{dt}
= \frac{dN}{d\eta} \!\cdot\! \frac{d\eta}{dt} = \frac1{a(t)} \!\cdot\! 
\Gamma_{\rm flat} \; .
\end{equation}
Therefore, any sufficiently long wave length, massless particle which emerges 
during accelerated expansion can exist forever, but very few emerge.

Massless fermions and vector gauge bosons are conformally invariant, and so
are conformally coupled scalars. These particles do not, by themselves, give
rise to interesting effects in cosmology. However, massless, minimally coupled 
scalars are not conformally invariant, nor are gravitons, which obey the same 
linearized equation of motion \cite{Lifshitz:1945du}. On de Sitter background 
(which means $\epsilon = 0$, $H$ is constant and $a(t) = e^{Ht}$) the occupation 
number for a scalar of wave vector $\vec{k}$ (or each polarization of graviton) 
is staggering,
\begin{equation}
N(t,\vec{k}) = \Bigl[ \frac{H a(t)}{2 \Vert \vec{k}\Vert}\Bigr]^2 \; . 
\label{occupation}
\end{equation}
In addition to the obvious exponential growth, note also that the occupation
number only becomes of order one at horizon crossing, $\Vert\vec{k} \Vert \equiv
k = H a(t)$. This is very important because it guarantees that cosmological 
particle production is an infrared effect which can be studied reliably using 
general relativity as a low energy effective field theory, without needing 
its unknown ultraviolet completion.

The explosive growth evident in expression (\ref{occupation}) is the origin of
two observables from the epoch of primordial inflation: the power spectra of
gravitons \cite{Starobinsky:1979ty} and scalars \cite{Mukhanov:1981xt}. In fact,
the occupation number (\ref{occupation}) can be expressed in terms of the tensor
power spectrum $\Delta^2_{h}(k)$, not just for de Sitter ($\epsilon = 0$) but
for general $\epsilon < 1$,
\begin{equation}
N(t,\vec{k}) \longrightarrow \frac{\pi \Delta^2_{h}(k)}{64 G k^2} \!\times\! 
a^2(t) \; . \label{genN}
\end{equation}
The power spectra are tree order quantum effects but the quanta from which they
derive must interact with themselves and with other particles. In certain cases
these interactions lead to time dependent effects which grow as more and more
particles are ripped out of the vacuum.

The choice of vacuum is so contentious in cosmological QFT that I had better
explain how expression (\ref{occupation}) was derived. The spatial integral of 
the Lagrangian density of a massless, minimally coupled scalar $\varphi(t,\vec{x})$ 
can be written, using Parseval's theorem, as a Fourier mode sum of independent 
harmonic oscillators $\widetilde{\varphi}(t,\vec{k})$,
\begin{equation}
\int \!\! d^3x \, a^3 \Bigl[ \frac12 \dot{\varphi}^2 - \frac1{2 a^2} \Vert 
\vec{\nabla} \varphi\Vert^2 \Bigr] = \int \!\! \frac{d^3k}{(2\pi)^3} \, a^3 
\Bigl[\frac12 \vert \widetilde{\varphi}\vert^2 - \frac{k^2}{2 a^2} \vert 
\widetilde{\varphi}\vert^2\Bigr] \; . \label{freeL}
\end{equation}
Each wave vector $\vec{k}$ corresponds to an independent harmonic oscillator
with time-dependent mass and frequency,
\begin{equation}
m(t) \longrightarrow \frac{d^3k}{(2\pi)^3} \!\times\! a^3(t) \qquad , \qquad
\omega(t,k) \longrightarrow \frac{k}{a(t)} \; . \label{SHO}
\end{equation}
At any instant this system is a harmonic oscillator, so we can define the 
instantaneous occupation number $N$ by writing the expectation value of its 
energy as $E = \omega (\frac12 + N)$. For the de Sitter geometry the mode
function $u(t,k)$ is simple and we find,
\begin{eqnarray}
\lefteqn{u(t,k) = \frac{H}{\sqrt{2 k^3}} \Bigl[1 - \frac{ik}{a H}\Bigr] 
\exp\Bigl[\frac{i k}{a H}\Bigr] = \frac{H}{\sqrt{2 k^3}} \Bigl[1 + i k \eta\Bigr]
e^{-ik \eta} } \nonumber \\
& & \hspace{4cm} \Longrightarrow \frac12 a^3 \Bigl[ \vert \dot{u}\vert^2 
+ \frac{k^2}{a^2} \vert u\vert^2\Bigr] = \frac{k}{a} \Bigl[\frac12 + \Bigl( 
\frac{a H}{2 k}\Bigr)^2\Bigr] \; . \qquad \label{BDvac}
\end{eqnarray}
The mode function (\ref{BDvac}) is known as Bunch-Davies vacuum
\cite{Chernikov:1968zm,Schomblond:1976xc,Bunch:1978yq}, and one can 
see from the occupation number (\ref{occupation}) that it corresponds to a 
state which was empty in the distant past when $a(t) \rightarrow 0$.

\subsection{Time Dependence in Cosmological QFT}

Accelerated expansion changes cosmological quantum field theory in profound
ways. First, it is nonsense to base the theory on asymptotic scattering 
experiments. At least classically, the universe began with a singularity at some 
finite time, and no one knows how it will end. Fixtures of flat space QFT such
as Euclideanization and defining ``the vacuum'' as ``the unique, normalizable 
energy eigenstate'' come to seem quaint. Instead of in-out matrix elements 
between states which were free at asymptotically early and late times, we must
become accustomed to computing true expectation values in the presence of 
states which are defined at some finite time. Finally, the explosive production 
of massless, minimally coupled scalars and gravitons evident in expression 
(\ref{occupation}) implies that we must expect time-dependent effects as more 
and more virtual particles emerge from the vacuum. 

Consider a self-interacting scalar field on de Sitter background,
\begin{equation}
\mathcal{L} = -\frac12 \partial_{\mu} \varphi \partial_{\nu} \varphi 
g^{\mu\nu} \sqrt{-g} - \frac14 \lambda \varphi^4 \sqrt{-g} \; .
\label{phi4Lag}
\end{equation}
The first dimensionally regulated computation I ever did on de Sitter
background was the expectation value of this model's stress-energy tensor
at 1 and 2-loop order \cite{Onemli:2002hr,Onemli:2004mb},
\begin{equation}
T_{\mu\nu} = \partial_{\mu} \varphi \partial_{\nu} \varphi - \frac12
g_{\mu\nu} g^{\rho\sigma} \partial_{\rho} \varphi \partial_{\sigma}
\varphi - \frac14 g_{\mu\nu} \lambda \varphi^4 \; . \label{Tmunu}
\end{equation}
Although the vacuum is not unique, most people work on Bunch-Davies
vacuum, which we saw from (\ref{BDvac}) was empty in the distant past. 
However, cosmological particle production means that we must release the 
state at some finite time which can be taken to be $t = 0$, at which 
point the de Sitter scale factor $a(t) = e^{H t}$ is unity. 

Because the state is homogeneous and isotropic, the expectation value 
takes the perfect fluid form,
\begin{equation}
\Bigl\langle \Omega \Bigl\vert T_{\mu\nu}(t,\vec{x}) \Bigr\vert \Omega 
\Bigr\rangle = u_{\mu} u_{\nu} \!\times\! \Bigl[\rho(t) \!+\! p(t)\Bigr] 
+ g_{\mu\nu} \!\times\! p(t) \qquad , \qquad u_{\mu} \equiv \delta^0_{~\mu} 
\; . \label{TVEV}
\end{equation}
The renormalized result consists of a ``simple'' part which grows with 
time, or remains constant, plus a ``complicated'' part that falls off 
exponentially \cite{Kahya:2009sz},
\begin{eqnarray}
\rho(t) &\!\!\! = \!\!\!& \frac{3 H^2}{8\pi G} + \frac{\lambda H^4}{(2\pi)^4}
\Biggl\{ 2 \ln^2(a) + \frac{13}{6} \ln(a) - \frac{43}{18} + \frac{\pi^2}{3}
\nonumber \\
& & \hspace{5.4cm} - \frac{3}{2 a^2} - 2 \sum_{n=4}^{\infty} \frac{(n\!+\!1)}{
n^2 a^n} \Biggr\} + O(\lambda^2) \; , \qquad \label{rho} \\
p(t) &\!\!\! = \!\!\!& -\frac{3 H^2}{8\pi G} + \frac{\lambda H^4}{(2\pi)^4}
\Biggl\{-2 \ln^2(a) - \frac{7}{2} \ln(a) + \frac{5}{3} - \frac{\pi^2}{3}
\nonumber \\
& & \hspace{4.4cm} + \frac{1}{2 a^2} - \frac23 \sum_{n=4}^{\infty} 
\frac{(n\!-\!3) (n\!+\!1)}{n^2 a^n} \Biggr\} + O(\lambda^2) \; . \qquad
\label{p} 
\end{eqnarray}
A nice check on accuracy is that (\ref{rho}-\ref{p}) obey conservation,
$\dot{\rho} = -3 (\rho + p)$. 

\subsection{Eliminating the Divergence with the Initial State}

In addition to being complicated, the infinite sums actually cause expressions
(\ref{rho}-\ref{p}) to diverge at $t=0$, for which $a = 1$. These divergences
have nothing to do with inflationary particle production. They derive instead 
from the initial state having been free, in spite of the interaction. Similar 
divergences would occur in flat space QFT if we had specified the state to be 
free vacuum at some finite time. The cure for these divergences is to correct 
the initial state. Note that we do not need the full state --- no one is ever 
going to find the full state wave functional for an interacting quantum field 
theory in $3+1$ dimensions. All we need is the order $\lambda$ correction; 
only this part of the initial state can affect our perturbative results 
(\ref{rho}-\ref{p}).

It is well to recall how a quartic interaction would change the ground state
of a simple harmonic oscillator (\ref{SHOLag}) in quantum mechanics,
\begin{equation}
H = \frac{p^2}{2m} + \frac{m \omega^2 q^2}{2} + \frac{\lambda q^4}{4} \; , \;
\vert \Omega \rangle = \vert 0 \rangle + \sum_{n=1}^{\infty} a_n \vert n \rangle
\Rightarrow a_n = -\frac{\lambda \langle n \vert q^4 \vert 0 \rangle}{4 n
\omega} + O(\lambda^2) \; .
\end{equation}
If we release the system at $t=0$ in a functional integral over configurations 
$q(t)$ then its wave function depends upon the position $q_0 \equiv q(t=0)$. The 
first order correction is,
\begin{equation}
\Omega_1(q_0) = -\frac{\lambda}{16 m^2 \omega^3} \Bigl[1 - 2 m \omega
q_0^2\Bigr]^2 \times \Omega_0(q_0) \; , \label{order1QM}
\end{equation}
where $\Omega_0(q_0)$ is the harmonic oscillator ground state (\ref{ground}).

If we release the QFT state at $t=0$ in the functional formalism then the
state wave functional will depend on $\varphi_0(\vec{x}) \equiv \varphi(t=0,\vec{x})$.
From the preceding quantum mechanical discussion we expect that the order
$\lambda$ correction to Bunch-Davies vacuum $\Omega_0[\varphi_0]$ will involve
zero, two and four powers of $\varphi_0$. Because the stress tensor operator 
(\ref{Tmunu}) has the form $\partial \varphi \partial \varphi + \lambda \varphi^4$, 
only the $\partial \varphi \partial \varphi$ part can couple with the order $\lambda$ 
state correction to contribute to the expectation value at order $\lambda$. If
this contribution is to depend upon the time $t$ at which the stress tensor is
evaluated, it must couple with the order $\lambda$ state correction involving
two powers of $\varphi_0$. Homogeneity and isotropy then imply a correction of 
the form, 
\begin{equation}
\Omega_1[ \varphi_0] = \frac{\lambda H}{2} \! \int \!\! \frac{d^3k}{(2\pi)^3} \, 
F\Bigl(\frac{k}{H}\Bigr) \Bigl\vert \widetilde{\varphi}_0(\vec{k}) \Bigr\vert^2 
\times \Omega_0[ \varphi_0] \; . \label{order1QFT}
\end{equation}
A straightforward calculation reveals that the kernel is \cite{Kahya:2009sz},
\begin{equation}
F(x) = \frac{i e^{2 i x}}{32 \pi^2 (1 \!+\! i x)^2} \Biggl\{ e^{-2i x} - x^3 \!
\int \!\! \frac{dz}{z^4} \, e^{-2i z} \Bigl[1 + 2 \ln\Bigl( \frac{z}{x}\Bigr)\Bigr]
\Biggr\} \; . \label{Fdef}
\end{equation}
And the corrected energy density and pressure are free of both divergences and
exponentially falling contributions,
\begin{eqnarray}
\rho_{\rm new} & \!\!\! = \!\!\! & \frac{3 H^2}{8\pi G} + \frac{\lambda H^4}{(2\pi)^4}
\Biggl\{2 \ln^2(a) + \frac{13}{6} \ln(a) - \frac{43}{18} + \frac{\pi^2}{3} \Biggr\} 
+ O(\lambda^2) \; , \label{rhonew} \\
p_{\rm new} & \!\!\! = \!\!\! & -\frac{3 H^2}{8\pi G} + \frac{\lambda H^4}{(2\pi)^4}
\Biggl\{-2 \ln^2(a) - \frac{7}{2} \ln(a) + \frac{5}{3} - \frac{\pi^2}{3} \Biggr\} 
+ O(\lambda^2) \; . \label{pnew}
\end{eqnarray}

\subsection{Can We Absorb the Logarithms?}

The factors of $\ln(a) = H t$ in expressions (\ref{rhonew}-\ref{pnew}) derive from
inflationary particle production. There are two such factors because the expectation
value of the $\varphi^4$ part of the stress tensor (\ref{Tmunu}) involves two 
coincident propagators, each of which contributes a $\ln(a)$ \cite{Vilenkin:1982wt,
Linde:1982uu,Starobinsky:1982ee},
\begin{equation}
\Bigl\langle \Omega_0 \Bigl\vert \varphi^2(t,\vec{x}) \Bigr\vert \Omega_0 \Bigr\rangle
= {\rm UV\ Divergence} + \frac{H^2}{4\pi^2} \ln(a) \; . \label{oldVEV}
\end{equation}
The physical way of understanding this is that inflationary particle production forces
the scalar further and further up its potential.

Higher loop corrections involve even more factors of $\ln(a)$. At order $\lambda^N$
there can be up to $2N$ factors of $\ln(a)$ \cite{Tsamis:2005hd}. Starobinsky has 
inferred a stochastic formalism \cite{Starobinsky:1986fx} which can be proven to
reproduce the leading logarithms at each order in perturbation theory 
\cite{Tsamis:2005hd}. In cases, such as this model (\ref{phi4Lag}), where a 
time-independent limit is approached, Starobinsky's technique can even be used to
find this limit \cite{Starobinsky:1994bd},
\begin{equation}
\Bigl\langle \Omega \Bigl\vert T_{\mu\nu}(t,\vec{x}) \Bigr\vert \Omega \Bigr\rangle
\longrightarrow \frac{3 H^4}{32 \pi^2} \, g_{\mu\nu} \; . \label{latetime} 
\end{equation}
The physical picture is that inflationary particle production pushes the scalar
up its potential until an equilibrium is reached with the classical downward force.
Of course that equilibrium corresponds to a particular state wave functional and,
were one to release the system in this state, there would be no time dependence.

Some people invoke the existence of an equilibrium state to argue that inflationary
particle production is somehow not real, or else does not induce time dependence
in QFT. This shows the same level of disingenuity as denying the inflationary origin
of primordial perturbations because one could obtain identical results by starting 
with a state which had exactly the desired pattern of correlations. However, the
anti-time dependence position becomes completely untenable when one considers 
models for which no static limit is approached. Although scalar quantum 
electrodynamics approaches a static limit \cite{Prokopec:2007ak}, a Yukawa-coupled
massless, minimally coupled scalar does not \cite{Miao:2006pn}. Nonlinear sigma
models also show large logarithms \cite{Tsamis:2005hd,Kitamoto:2010et,
Kitamoto:2011yx,Kitamoto:2018dek}, and some of these models do not approach a static
limit \cite{Miao:2021gic,Woodard:2023rqo}.

Nonlinear sigma models are interesting in that capturing their large logarithms
requires a variant of the renormalization group in addition to a variant of
Starobinsky's stochastic formalism \cite{Miao:2021gic,Woodard:2023rqo}. This is
because their interactions involve derivatives, unlike the $\varphi^4$ interaction
of the scalar model (\ref{phi4Lag}). Quantum gravity also involves derivative
interactions and its large inflationary logarithms \cite{Miao:2006gj,Glavan:2013jca,
Wang:2014tza,Glavan:2021adm,Park:2015kua,Tan:2021lza,Tan:2022xpn} are not completely
captured by Starobinsky's stochastic formalism \cite{Miao:2008sp}. I suspect that 
combining Starobinsky's formalism with a variant of the renormalization group will 
suffice \cite{Glavan:2021adm}, as it did for nonlinear sigma models, but that 
remains to be proven. It is not known if the large logarithms of quantum gravity 
add up to approach a static limit.

\section{Conclusions}

Steven Weinberg dominated high energy particle theory for decades and inspired
generations of physicists. I was fortunate to be among them. Two of the things I 
most admired about Weinberg were his refusal to accept dogma and his willingness 
to take risks. In addition to sharing some reminiscences, I have here discussed 
a topic which illustrates both of those characteristics: the role of state wave 
functionals in quantum field theory.

State wave functionals are the correct way to derive the $i\epsilon$ of the 
Feynman propagator from the flat space functional formalism. They assume a larger 
role in cosmological QFT because one must specify states at finite time, and because 
inflationary particle production sometimes injects secular growth into expectation 
values, the potential for which Weinberg discovered in the scalar power spectrum
\cite{Weinberg:2006ac}. Perturbative corrections to free vacuum are necessary to 
remove divergences on the initial value surface \cite{Kahya:2009sz}, which leaves 
large logarithms from inflationary particle production. Weinberg realized the 
importance of resumming these logarithms \cite{Weinberg:2006ac}. In some cases, 
they can be summed to produce a static limit which can be subsumed into a highly 
nonlinear state wave functional \cite{Starobinsky:1994bd}. For other models, time 
dependence persists forever \cite{Miao:2021gic,Woodard:2023rqo} and one is tempted
to wonder if some otherwise curious features of late time cosmology might be
explained as residual QFT effects \cite{Tsamis:2011ep,Woodard:2018gfj}. Wherever 
he is now, Big Steve would love that.

\vskip 0.5cm

\centerline{\bf Acknowledgements}

I am grateful for conversations on this subject with S. Deser, D. Glavan, E. O.
Kahya, S. P. Miao, V. K. Onemli, T. Prokopec and N. C. Tsamis. This work was 
supported by NSF grant PHY-2207514 and by the Institute for Fundamental Theory 
at the University of Florida.

\end{document}